\documentclass{article}

\usepackage{microtype}
\usepackage{graphicx}
\usepackage{subcaption}
\usepackage{booktabs} 
\usepackage{multirow}
\usepackage{bm} 

\usepackage{hyperref}

\usepackage[preprint]{icml2026}

\usepackage{amsmath}
\usepackage{amssymb}
\usepackage{mathtools}
\usepackage{amsthm}

\usepackage[capitalize,noabbrev]{cleveref}

\theoremstyle{plain}

\theoremstyle{definition}

\theoremstyle{remark}

\usepackage[textsize=tiny]{todonotes}

\icmltitlerunning{Functional Subspace Watermarking for Large Language Models}

\begin{document}

\twocolumn[
  \icmltitle{Functional Subspace Watermarking for Large Language Models}



  \icmlsetsymbol{equal}{*}

  \begin{icmlauthorlist}
    \icmlauthor{Zikang Ding}{equal,yyy,comp}
    \icmlauthor{Junhao Li}{equal,sch}
    \icmlauthor{Suling Wu}{yyy}
    \icmlauthor{Junchi Yao}{yyy,comp}
    \icmlauthor{Hongbo Liu}{yyy}
    \icmlauthor{Lijie Hu}{comp}

  \end{icmlauthorlist}

  \icmlaffiliation{yyy}{University of Electronic Science and Technology of China, Chengdu, China}
  \icmlaffiliation{comp}{Mohamed bin Zayed University of Artificial Intelligence, Abu Dhabi, United Arab Emirates}
  \icmlaffiliation{sch}{South China University of Technology, Guangzhou, China}

  \icmlcorrespondingauthor{Lijie Hu}{lijie.hu@mbzuai.ac.ae}
  

  \icmlkeywords{Machine Learning, ICML}

  \vskip 0.3in
]



\printAffiliationsAndNotice{}  

\begin{abstract}
Model watermarking utilizes internal representations to protect the ownership of large language models (LLMs). However, these features inevitably undergo complex distortions during realistic model modifications such as fine-tuning, quantization, or knowledge distillation, making reliable extraction extremely challenging. Despite extensive research on model-side watermarking, existing methods still lack sufficient robustness against parameter-level perturbations. To address this gap, we propose \texttt{\textbf{Functional Subspace Watermarking (FSW)}}, a framework that anchors ownership signals into a low-dimensional functional backbone. Specifically, we first solve a generalized eigenvalue problem to extract a stable functional subspace for watermark injection, while introducing an adaptive spectral truncation strategy to achieve an optimal balance between robustness and model utility. Furthermore, a vector consistency constraint is incorporated to ensure that watermark injection does not compromise the original semantic performance. Extensive experiments across various LLM architectures and datasets demonstrate that our method achieves superior detection accuracy and statistical verifiability under multiple model attacks, maintaining robustness that outperforms existing state-of-the-art (SOTA) methods.
\end{abstract}


\section{Introduction}
\begin{figure}[ht]
    \centering
    \includegraphics[width=\linewidth]{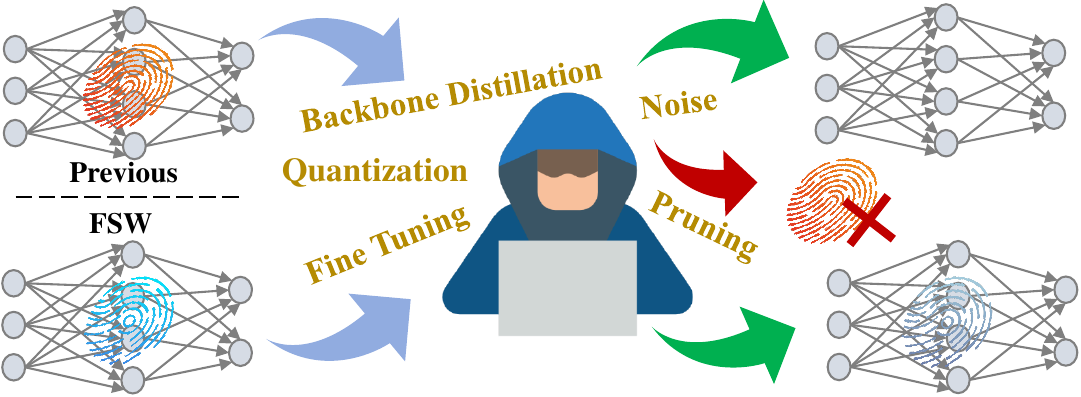} 
    \caption{Conceptual comparison of watermark robustness under common model-side attacks. \textbf{Top (Previous):} Conventional internal watermarks are vulnerable to post-hoc modifications like fine-tuning, quantization, pruning, and distillation, which often result in signal erasure. \textbf{Bottom (\texttt{\textbf{FSW}}):} Our proposed framework anchors ownership signals into a stable functional backbone, ensuring that the watermark remains detectable even after significant parameter-level perturbations.}
    \label{fig:attack_overview}
\end{figure}

Large language models (LLMs) and multimodal LLMs have advanced rapidly in recent years, driven in part by the widespread release of capable open weight models such as Llama \citep{meta2025llama4_multimodal_intelligence}, Mixtral \citep{mistral2025mistral3}, Qwen \citep{yang2025qwen3technicalreport}, and Gemma \citep{gemmateam2025gemma3technicalreport}. As these models become easier to download, fine tune, and redistribute, questions of model ownership and provenance become increasingly important. In practice, model developers face realistic threats including checkpoint leakage, unauthorized redistribution, and model extraction through supervised querying and knowledge distillation, all of which can transfer capabilities while obscuring attribution.

Existing provenance and ownership protections for LLMs mainly fall into two categories. The first is content watermarking, which embeds statistically detectable signals into generated texts for provenance tracing without requiring access to model parameters \citep{kirchenbauer2023watermark,dathathri2024scalable,zhang2024remark}. The second is model watermarking, which aims to embed ownership information into model weights or internal representations so that a suspect model can be verified in an ownership dispute \citep{li2023watermarking,sander2024watermarking,adi2018turning}. While these directions are complementary, both face important limitations in adversarial settings. Content watermarking is primarily designed to attribute outputs and is vulnerable to strong extraction pipelines, especially when an adversary distills a student model that reproduces capabilities without preserving the original output level signal (As shown in Figure \ref{fig:attack_overview}). Model internal watermarking for LLMs remains relatively under explored, which is still a central challenge.

To address this gap, we propose \texttt{\textbf{FSW}}, which embeds ownership signals into the functional backbone of a model. We identify a low-dimensional latent subspace that is critical to task performance and remains stable under a broad class of compression-related transformations, enabling robustness to common model modifications. Based on this subspace, we formulate a generalized eigenvalue framework that explicitly decouples task-critical sensitivity from compression-induced variance, allowing the extraction of a stable watermark carrier. Furthermore, we introduce a spectral truncation strategy that selects a specific range to achieve an optimal balance between robustness and model utility. Experimental results show that this targeted selection maintains high watermark verifiability while strictly preserving the original model performance under various model modifications, including fine-tuning, quantization, and knowledge distillation.

This paper makes the following key contributions:

\begin{enumerate}

    \item We propose the \texttt{\textbf{FSW}} framework, which embeds watermarks by locating a functional backbone in the model that is both critical and stable. This method uses a generalized eigenvalue problem to extract carrier directions, effectively defending against various compression transformations and distillation attacks.

    \item We design an orthogonal-key-based multi-bit watermark verification scheme, which transforms the ownership verification process into a rigorous statistical test. Without affecting the model's original generation performance, this scheme achieves reliable watermark injection and accurate verification.

    \item We conduct extensive experiments on multiple model architectures and datasets to validate the robustness of \texttt{\textbf{FSW}} against fine-tuning, quantization, and knowledge distillation. The results demonstrate that, while maintaining model performance, the framework can complete ownership verification at an extremely low false positive rate.

\end{enumerate}
\begin{figure*}[t]
    \centering
    \includegraphics[width=1\linewidth]{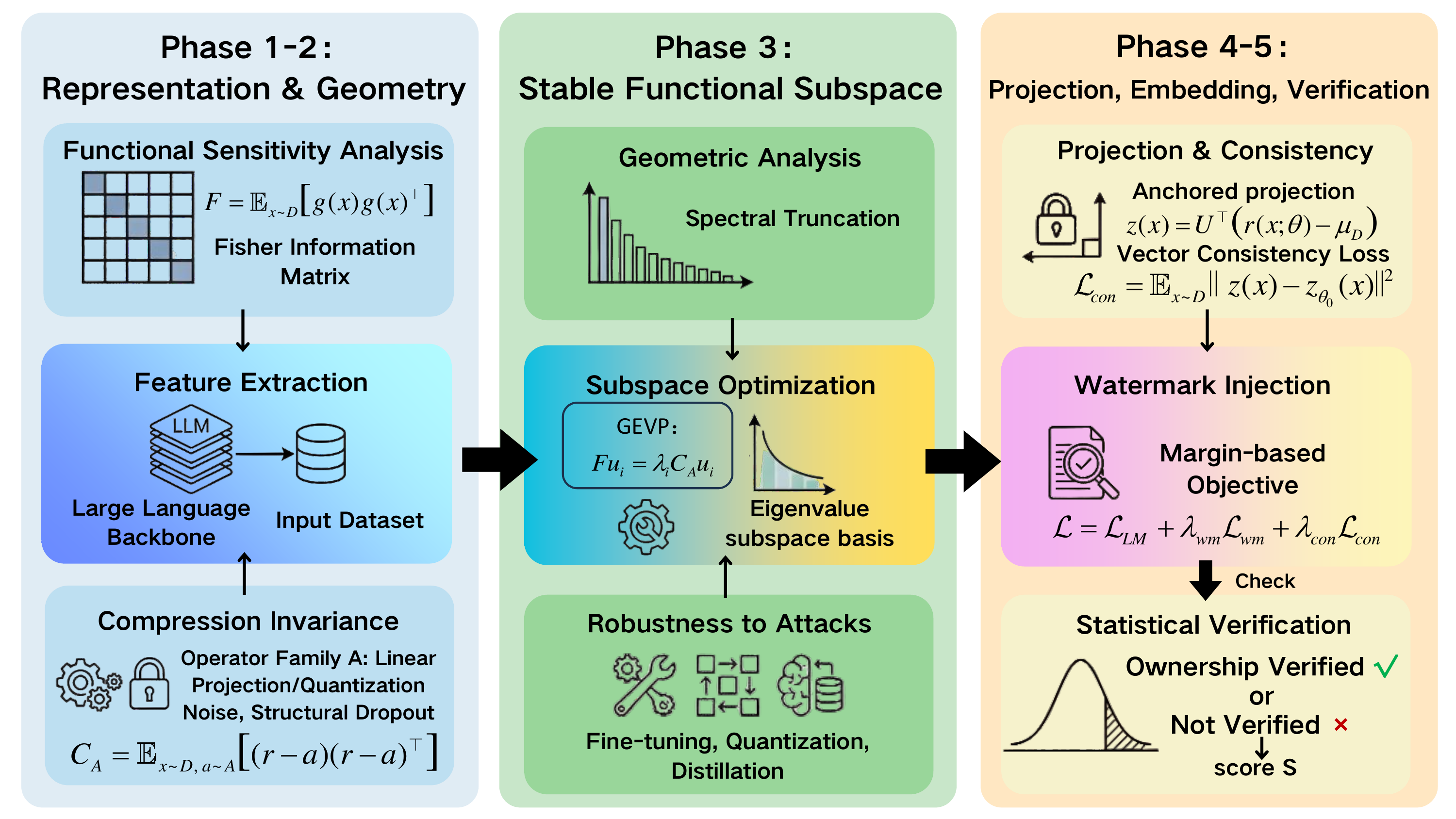}
    \caption{Overall framework of FSW.}
    \label{fig:framework}
\end{figure*}
\section{Related Work}
\subsection{Content Watermarking}
Content watermarking embeds statistically detectable signals into generated texts to enable provenance tracing without requiring access to model parameters. Many approaches operate at generation time by steering token selection with a secret key so that the resulting text exhibits a measurable signature. A representative instance is the greenlist-based design of \citet{kirchenbauer2023watermark}, and subsequent systems improve robustness, payload, and sampling efficiency across broader settings \citep{zhang2024remark,mao2025watermarking,niess2025ensemble}. Moving toward deployment, \citet{dathathri2024scalable} demonstrate watermarking at practical scale for identifying LLM outputs, showing that such detectors can be integrated into real inference pipelines.

Recent work further explores robustness against post generation edits and stronger verification. \citet{dabiriaghdam-wang-2025-simmark} is notable for shifting the carrier from token statistics to sentence level semantic similarity, improving robustness to paraphrasing while remaining compatible with API-only access. Another direction targets stronger payload under paraphrasing by explicitly leveraging LLM-based paraphrasers to encode and recover multi-bit messages \citep{xurobust}. Orthogonal to these, ensemble-style schemes combine multiple watermark signals to improve detectability and resilience across diverse attack patterns \citep{niess2025ensemble}, while in-context watermarking explores watermark injection and detection through prompt-time mechanisms rather than modifying the underlying model \citep{liu2025context}. On the theory side, distribution adaptive frameworks aim to ground watermark design in statistical guarantees under varying output distributions \citep{hetheoretically}.

\subsection{Model Watermarking}
Model watermarking aims to embed ownership information into model parameters or internal representations so that suspect models can be verified in ownership disputes, even after common post-hoc modifications. Classic approaches in deep neural networks include trigger-set and backdoor based ownership proofs \citep{adi2018turning}, end-to-end watermark embedding frameworks \citep{darvish2019deepsigns}, and covert, robust white-box watermarking designs \citep{wang2021riga}.

For LLMs, internal watermarking remains relatively under-explored. Existing efforts include embedding signals through weight quantization \citep{li2023watermarking}, leveraging ``radioactive'' training traces for downstream attribution \citep{sander2024watermarking}, and SEAL, which anchoring ownership signals in latent subspaces \citep{dai2025seal}. Related black-box ownership verification via fingerprinting has also been studied under model merging, for example MERGEPRINT \citep{yamabe2025mergeprint}. Despite progress, robustness under strong extraction, especially knowledge distillation, remains a key challenge. This leaves an important gap for model-side watermarking methods designed explicitly for distillation settings. To address this gap, we propose \texttt{\textbf{FSW}}, which embeds ownership information in functional subspaces that are important for the task and remain stable under compression and distillation, improving verifiability under distillation threats.

\section{Threat Model}

We consider a model ownership protection scenario where a model owner releases a watermarked LLM and subsequently identifies a suspect model of disputed provenance. The adversary's objective is to remove or invalidate the embedded signature while maintaining the model's practical utility for downstream tasks \citep{wang2021riga, xu2024instructional}.

\paragraph{Adversary Capabilities and Modifications.}
Following the taxonomy of model lineage auditing \citep{shao2025sok}, we assume a strong adversary with white-box access to the model parameters \citep{pasquini2025llmmap}. The adversary can perform various post-hoc modifications to evade detection, including supervised fine-tuning (SFT), parameter-efficient adaptation (e.g., LoRA), and model compression techniques such as weight pruning and quantization \citep{xu2024instructional, yan2023rethinking}. Crucially, the adversary is permitted to optimize the modified model to restore task-level performance, ensuring it remains competitive in terms of perplexity and downstream accuracy \citep{xu2024instructional}.

\paragraph{Functional Preservation Constraint.}
Consistent with the requirements for robust internal watermarking \citep{yao2024promptcare}, we restrict the adversary to attacks that preserve the model's internal functional backbone representations. This constraint reflects realistic deployment settings where adversaries prioritize the structural integrity and reliability of the original model over re-engineering its internal logic from scratch. Accordingly, our guarantees do not extend to fully unconstrained distillation where the student model is free to arbitrarily reparameterize its latent space \citep{shao2025sok, xu2024instructional}.

\paragraph{Verification Protocol.}
We assume the defender has white-box access to the suspect model for ownership verification \citep{wang2021riga}. This allows for the inspection of model parameters or intermediate activations to establish proof of ownership \citep{li2025differentiation}. By analyzing representations within the identified functional subspace, the defender can recover embedded signatures even after significant parameter-level perturbations or structural obfuscations \citep{yan2023rethinking}.

\section{Methodology}
We propose \texttt{\textbf{FSW}}, which embeds ownership signatures into a stable \emph{functional backbone} by solving a Generalized Eigenvalue Problem (GEVP). This backbone is optimized for task-criticality and compression resilience to ensure watermark persistence during distillation or quantization. The implementation follows a structured four-phase pipeline: 

\subsection{Phase 1: Representation Extraction}
Let $f_\theta$ denote an autoregressive language model with $L$ layers and hidden dimension $d$. Given an input sequence $x = (x_1, \dots, x_T)$, we extract the representation from a fixed intermediate layer $\ell$:
\begin{equation}
    r(x; \theta) = H^{(\ell)}_{T} \in \mathbb{R}^{d},
\end{equation}
where $H^{(\ell)}_{T}$ is the hidden state of the last token. This state serves as the information bottleneck for next-token prediction, making it the optimal carrier for semantic-level watermarking.
\begin{table*}[t]
\centering
\small
\setlength{\tabcolsep}{8pt} 
\begin{tabular*}{\textwidth}{@{\extracolsep{\fill}} ll cccc ccc}
\toprule
\multirow{2.5}{*}{\textbf{Model}} & \multirow{2.5}{*}{\textbf{Method}} & \multicolumn{4}{c}{\textbf{Functional Preservation}} & \multicolumn{3}{c}{\textbf{Detectability}} \\
\cmidrule(lr){3-6} \cmidrule(lr){7-9}
& & PPL $\downarrow$ & $\Delta$PPL $\downarrow$ & HellaSwag $\uparrow$ & ARC-E $\uparrow$ & Det. Score $\uparrow$ & Bit Acc $\uparrow$ & AUC $\uparrow$ \\
\midrule

\multirow{2}{*}{\textbf{LLaMA-2-7B-hf}} 
& Clean FT & 5.30 & 0.00 & 66.40 & 71.60 & 0.47 & -- & -- \\
& \textbf{FSW (Ours)} & 5.91 & +0.61 & 64.80 & 71.00 & \textbf{6.09} & 100\% & 0.894 \\
\midrule

\multirow{2}{*}{\textbf{LLaMA-3-8B}} 
& Clean FT & 5.83 & 0.00 & 68.60 & 79.60 & -0.03 & -- & -- \\
& \textbf{FSW (Ours)} & 6.20 & +0.37 & 67.80 & 75.80 & \textbf{4.00} & 100\% & 0.899 \\
\midrule

\multirow{2}{*}{\textbf{Qwen2.5-7B}} 
& Clean FT & 6.35 & 0.00 & 67.00 & 75.80 & 1.37 & -- & -- \\
& \textbf{FSW (Ours)} & 7.24 & +0.89 & 66.80 & 73.40 & \textbf{5.09} & 100\% & 0.977 \\
\midrule

\multirow{2}{*}{\textbf{Mistral-7B-v0.3}} 
& Clean FT & 16.68 & 0.00 & 68.60 & 77.00 & -0.22 & -- & -- \\
& \textbf{FSW (Ours)} & 17.25 & +0.57 & 69.00 & 77.00 & \textbf{3.84} & 100\% & 0.999 \\
\midrule

\multirow{2}{*}{\textbf{DeepSeek-7B-Chat}} 
& Clean FT & 59.52 & 0.00 & 67.40 & 72.20 & -0.09 & -- & -- \\
& \textbf{FSW (Ours)} & 77.31 & +17.79 & 60.40 & 66.20 & \textbf{3.75} & 100\% & 1.000 \\
\bottomrule
\end{tabular*}
\caption{Performance comparison of Functional Subspace Watermarking (FSW) across diverse LLM backbones. Our method achieves near-perfect bit accuracy and high detection scores while maintaining competitive functional utility.}
\label{tab:1}
\end{table*}

\begin{table*}[t]
\centering
\small
\setlength{\tabcolsep}{8pt} 
\begin{tabular*}{\textwidth}{@{\extracolsep{\fill}} ll ccc cc}
\toprule
\multirow{2.5}{*}{\textbf{Model}} & \multirow{2.5}{*}{\textbf{Attack}} & \multicolumn{3}{c}{\textbf{Watermark Reliability}} & \multicolumn{2}{c}{\textbf{Model Quality}} \\
\cmidrule(lr){3-5} \cmidrule(lr){6-7}
& & Det. Score $\uparrow$ & Ret. $\uparrow$ & Bit Acc $\uparrow$ & PPL $\downarrow$ & $\Delta$PPL $\downarrow$ \\
\midrule

\multirow{6}{*}{\textbf{LLaMA-2-7B-hf}} 
& Baseline & 6.0938 & 100.00\% & 100.00\% & 5.91 & +0.00  \\
& Backbone Distillation & 5.8438 & 95.90\% & 87.50\% & 5.91 & +0.00  \\
& LoRA FT & 6.7812 & 111.28\% & 87.50\% & 5.58 & -0.33  \\
& Noise & 5.9062 & 96.92\% & 100.00\% & 5.92 & +0.01  \\
& Pruning & 6.7188 & 110.26\% & 87.50\% & 6.13 & +0.22  \\
& Quantization & 6.5625 & 107.69\% & 100.00\% & 6.43 & +0.51  \\
\midrule

\multirow{6}{*}{\textbf{Meta-LLaMA-3-8B}} 
& Baseline & 4.0000 & 100.00\% & 100.00\% & 6.20 & +0.00  \\
& Backbone Distillation & 3.1719 & 79.30\% & 62.50\% & 6.12 & -0.08  \\
& LoRA FT & 4.2188 & 105.47\% & 87.50\% & 6.10 & -0.10  \\
& Noise & 4.1250 & 103.12\% & 100.00\% & 6.21 & +0.01  \\
& Pruning & 3.2812 & 82.03\% & 87.50\% & 6.27 & +0.07  \\
& Quantization & 4.2500 & 106.25\% & 87.50\% & 6.65 & +0.45  \\
\midrule

\multirow{6}{*}{\textbf{Qwen2.5-7B}} 
& Baseline & 5.0938 & 100.00\% & 100.00\% & 7.24 & +0.00  \\
& Backbone Distillation & 3.1406 & 61.66\% & 75.00\% & 7.66 & +0.42  \\
& LoRA FT & 3.4688 & 68.10\% & 87.50\% & 6.75 & -0.49  \\
& Noise & 4.6875 & 92.02\% & 100.00\% & 7.26 & +0.02  \\
& Pruning & 4.8438 & 95.09\% & 100.00\% & 7.32 & +0.08  \\
& Quantization & 5.2500 & 103.07\% & 100.00\% & 8.22 & +0.98  \\
\midrule

\multirow{6}{*}{\textbf{Mistral-7B-v0.3}} 
& Baseline & 3.8438 & 100.00\% & 100.00\% & 17.25 & +0.00  \\
& Backbone Distillation & 3.2344 & 84.15\% & 75.00\% & 6.59 & -10.66  \\
& LoRA FT & 3.7812 & 98.37\% & 100.00\% & 16.92 & -0.33  \\
& Noise & 3.7656 & 97.97\% & 100.00\% & 17.19 & -0.05  \\
& Pruning & 3.3281 & 86.59\% & 100.00\% & 17.58 & +0.34  \\
& Quantization & 3.7969 & 98.78\% & 100.00\% & 18.99 & +1.75  \\
\midrule

\multirow{6}{*}{\textbf{DeepSeek-LLM-7B-Chat}} 
& Baseline & 3.7500 & 100.00\% & 100.00\% & 77.31 & +0.00  \\
& Backbone Distillation & 1.6562 & 44.17\% & 100.00\% & 61.39 & -15.92  \\
& LoRA FT & 0.6602 & 17.60\% & 100.00\% & 62.31 & -15.00  \\
& Noise & 3.7344 & 99.58\% & 100.00\% & 77.28 & -0.03  \\
& Pruning & 3.4844 & 92.92\% & 100.00\% & 81.79 & +4.48  \\
& Quantization & 3.7188 & 99.17\% & 100.00\% & 92.76 & +15.45  \\
\bottomrule
\end{tabular*}
\caption{Robustness analysis of FSW under various model attacks. The results across multiple LLM backbones show that ownership signals remain recoverable even after significant parameter-level modifications.}
\label{tab:robustness}
\end{table*}

\subsection{Phase 2: Geometry Analysis}
To construct a stable watermark injection subspace, we analyzed the characteristics of the subspace from two aspects:
\paragraph{Functional Sensitivity.} 
We quantify the task relevance of representation directions using the Fisher information matrix $F$, estimated on a calibration set drawn from the target distribution $\mathcal{D}$:
\begin{equation}
    F = \mathbb{E}_{x \sim \mathcal{D}} \left[ g(x) g(x)^\top \right] \in \mathbb{R}^{d \times d},
\end{equation}
where $g(x) = \nabla_{r} L_{\mathrm{LM}}(x; \theta)$. Directions with large quadratic forms $u^\top F u$ encode the model's core knowledge. Perturbing these directions significantly alters the output distribution, making them inherently resistant to pruning-based removal \citep{yan2023rethinking}.

\paragraph{Invariance to Compression Operators.} 
Rather than overfitting to a specific student architecture, we model knowledge distillation as a process of information compression. We define a family of Compression Operators $\mathcal{A}$ that approximate shared invariants, including:
\begin{itemize}
    \item \textit{Linear Projection:} $a(r) = W_{\mathrm{low}} r$, simulating capacity bottlenecks.
    \item \textit{Quantization Noise:} $a(r) = r + \epsilon,\ \epsilon \sim \mathcal{N}(0, \sigma^2 I)$, simulating precision loss.
    \item \textit{Structural Dropout:} $a(r) = r \odot m$, simulating feature sparsification.
\end{itemize}
We define the transformation-invariant matrix $C_{\mathcal{A}}$ as the expected variance caused by these operators:
\begin{equation}
    C_{\mathcal{A}} = \mathbb{E}_{x \sim \mathcal{D},\, a \sim \mathcal{A}} \left[ (r(x) - a(r(x))) (r(x) - a(r(x)))^\top \right].
\end{equation}
Minimizing $u^\top C_{\mathcal{A}} u$ encourages the watermark to reside in directions that remain stable under compression-equivalent transformations \citep{yao2024promptcare}.

\subsection{Phase 3: Subspace Construction \& Optimization}
We seek a subspace $U^\star \in \mathbb{R}^{d \times k}$ that maximizes the signal-to-noise ratio between functional sensitivity and compression-induced variance. This objective is formulated as the GEVP:
\begin{equation}
    F u_i = \lambda_i C_{\mathcal{A}} u_i,
\end{equation}
where the generalized eigenvalue $\lambda_i$ represents the robustness score of the $i$-th direction.

\paragraph{Spectral Truncation.} 
To address the issue that using top-k feature vectors leads to a catastrophic decrease in utility, while using bottom feature vectors results in easy removal, we select the optimal point based on the following criteria:
\begin{equation}
    \mathcal{I} = \{ i \mid \tau_{\mathrm{lower}} \cdot \lambda_1 \le \lambda_i \le \tau_{\mathrm{upper}} \cdot \lambda_1 \}.
\end{equation}
Thresholds $\tau_{\mathrm{lower}}, \tau_{\mathrm{upper}}$ are determined via validation criteria ensuring functional integrity. The top-$k$ indices from $\mathcal{I}$ form the functional backbone subspace $U^\star$.

\subsection{Phase 4: Projection and Consistency}
We introduce an anchored projection to define a stable coordinate system:
\begin{equation}
    z(x) = U^{\star\top} (r(x; \theta) - \mu_{\mathcal{D}}) \in \mathbb{R}^{k},
\end{equation}
where $\mu_{\mathcal{D}}$ is the global mean representation from the frozen model $f_{\theta_0}$. To preserve behavior, we impose a {Vector Consistency constraint:
\begin{equation}
    L_{\mathrm{con}} = \mathbb{E}_{x \sim \mathcal{D}} \| z(x) - z_{\theta_0}(x) \|_2^2.
\end{equation}

\subsection{Phase 5: Watermark Embedding and Verification}

The model owner generates a challenge set $\mathcal{C}$ and samples $M$ mutually orthogonal secret key vectors $\{b_1, \dots, b_M\}$ with $b_j \in \mathbb{R}^k$, where $b_i^\top b_j \approx 0$ for $i \neq j$. Binary messages are encoded into target signs $y_j \in \{-1, 1\}$.

\paragraph{Embedding Loss.} 
We inject the watermark into the functional backbone using a margin-based objective:
\begin{equation}
    L_{\mathrm{wm}} = \mathbb{E}_{x \sim \mathcal{C}} \sum_{j=1}^M \max\left( 0, \gamma - y_j \frac{b_j^\top z(x)}{\|b_j\|_2} \right),
\end{equation}
where $\gamma > 0$ is the target margin. The watermarked model is obtained by minimizing the total objective:
\begin{equation}
    L = L_{\mathrm{LM}} + \lambda_{\mathrm{wm}} L_{\mathrm{wm}} + \lambda_{\mathrm{con}} L_{\mathrm{con}}.
\end{equation}

\paragraph{Statistical Verification.} 
Ownership is established by computing the aggregated detection score $S$:
\begin{equation}
    S = \frac{1}{|\mathcal{C}| M} \sum_{x \in \mathcal{C}} \sum_{j=1}^M y_j \frac{b_j^\top z(x)}{\|b_j\|_2}.
\end{equation}
Under the null hypothesis $H_0$, $S \sim \mathcal{N}(0, \sigma_0^2)$. The False Positive Rate (FPR) for a threshold $T$ is bounded by:
\begin{equation}
    \text{FPR} = \frac{1}{2} \mathrm{erfc} \left( \frac{T}{\sqrt{2}\sigma_0} \right).
\end{equation}

\paragraph{Multi-bit Decoding.} 
For message recovery, per-bit statistics are computed as $S_j = \frac{1}{|\mathcal{C}|} \sum_{x \in \mathcal{C}} \frac{b_j^\top z(x)}{\|b_j\|_2}$. The original bits are recovered via $\hat y_j = \mathrm{sign}(S_j)$ and decoded through Error-Correcting Codes (ECC). 

The complete procedure for watermark generation, embedding, and statistical verification is summarized in Algorithm~\ref{algorithm} in the Appendix \ref{appendix a}.

\section{Experiments}

\subsection{Experimental Setting}
\begin{figure}[t]
    \centering
    \includegraphics[width=\linewidth]{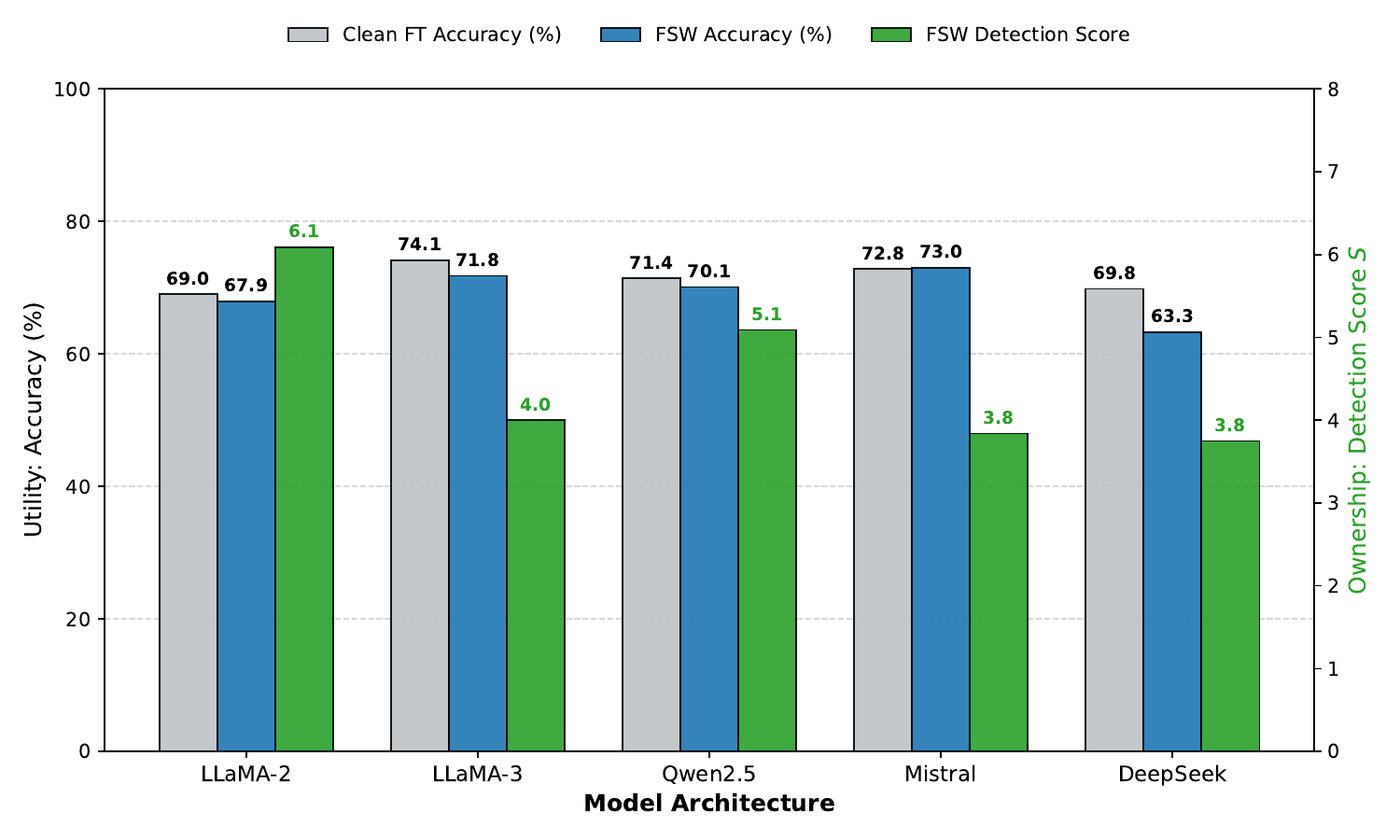}
    \caption{Functional Preservation and Ownership Detectability.} 
    \label{fig:3}
\end{figure}

\paragraph{Models and Datasets.} 

To evaluate the effectiveness and generality of FSW across diverse architectures, we select representative autoregressive language models including LLaMA-2-7B-hf \citep{touvron2023llama}, Meta-LLaMA-3-8B \citep{roque2024evolution}, Qwen2.5-7B \citep{team2024qwen2}, Mistral-7B-v0.3 \citep{jiang2023mistral7b}, and DeepSeek-LLM-7B-Chat \citep{bi2024deepseek}. We employ a multi-stage data strategy: WikiText-2 \citep{liu2025evolution} is utilized for estimating the Fisher matrix $F$ and the compression invariance matrix $C_{\mathcal{A}}$; C4 \citep{zhu2023multimodal} is used for watermark embedding fine-tuning to improve distributional robustness. Downstream benchmarks including HellaSwag \citep{zellers2019hellaswag} and ARC \citep{clark2018think} are used to verify that watermark embedding does not degrade task-level capabilities. Detailed selection principles and dataset configurations are provided in Appendix~\ref{app:exp_details}.

\paragraph{Baseline.} 
To evaluate the superiority of FSW, we compared it with several state-of-the-art (SOTA) baseline methods and technical variants: (1) Clean FT ;(2) EmMark \citep{zhang2024emmark}; (3) Weighted Quantization \citep{li2023watermarking}; (4) Naive Top-k: a variant of our framework that selects the k directions with the highest eigenvalues $\lambda_i$ from GEVP without applying an adaptive spectral truncation strategy.

\paragraph{Implementation Details.} 
We extract the hidden representation of the last token from an intermediate layer $\ell$ as the representation vector. The invariance matrix $C_{\mathcal{A}}$ is estimated using a family of compression operators $\mathcal{A}$ simulating capacity bottlenecks (Linear Projection), precision loss (Quantization Noise), and feature sparsification (Structural Dropout). Subspace dimension $k$ and spectral truncation thresholds are selected via validation set perplexity constraints to balance robustness and utility. Complete hardware and training configurations are detailed in Appendix~\ref{Implementation Details}.
\begin{table}[t]
\centering
\small
\setlength{\tabcolsep}{6pt} 
\begin{tabular}{l cc cc}
\toprule
\multirow{2}{*}{\textbf{Bits}} & \multicolumn{2}{c}{\textbf{No ECC (\%)}} & \multicolumn{2}{c}{\textbf{Hamming (7,4) (\%)}} \\
\cmidrule(lr){2-3} \cmidrule(lr){4-5}
& Bit Acc $\uparrow$ & Msg Acc $\uparrow$ & Bit Acc $\uparrow$ & Msg Acc $\uparrow$ \\
\midrule
4  & 100.0 & 100.0 & 0.0   & 0.0   \\
8  & 87.5  & 87.5  & \textbf{100.0} & \textbf{100.0} \\
12 & 100.0 & 100.0 & 100.0 & 100.0 \\
16 & 100.0 & 100.0 & 93.8  & 100.0 \\
32 & 100.0 & 100.0 & 100.0 & 100.0 \\
64 & 100.0 & 100.0 & 100.0 & 100.0 \\
\bottomrule
\end{tabular}
\caption{Impact of Error-Correcting Codes (ECC) on message decoding accuracy (Llama-2-7B-hf). The results demonstrate that \texttt{Hamming (7,4)} successfully eliminates residual errors at the 8-bit critical point, ensuring irrefutable ownership verification.}
\label{tab:ecc_analysis}
\end{table}

\subsection{Main result}
\noindent{\textbf{Functional Performance.}}
Our experimental results demonstrate that FSW achieves exceptional functional preservation across diverse LLM architectures. As illustrated in Figure~\ref{fig:3}, the side-by-side comparison of downstream task accuracy between the Clean FT baseline (light grey bars) and FSW (blue bars) reveals minimal performance degradation. For instance, the PPL increment for Meta-LLaMA-3-8B is only $0.37$, while maintaining a high average accuracy of $71.8\%$ . Simultaneously, the green bars represent the high-magnitude detection scores $S$, showing a distinct separation from the null hypothesis across all backbones. This visualized method profile validates the efficacy of our spectral truncation strategy: by avoiding modifications to highly sensitive representation directions critical for task performance, FSW successfully embeds watermarks into a carrier space that is both robust and non-intrusive to the model's original capabilities. 

Furthermore, the generalization capability of FSW across different training corpora is highlighted in Table \ref{tab:dataset_generalization} in Appendix \ref{appendix c}. Even when the embedding fine-tuning is shifted from the high-quality WikiText-2 dataset to the massive C4 corpus, the framework maintains consistent functional utility and robust detectability on the Meta-LLaMA-3-8B backbone. For instance, while using the C4 dataset slightly increases the perplexity ($\Delta\text{PPL}=+0.90$), it yields a perfect bit accuracy of $100\%$ and a near-ideal AUC of $1.000$ . This suggests that the functional backbone identified by FSW captures fundamental task-relevant directions that are inherent to the model's architecture rather than being overfitted to a specific dataset, ensuring reliable ownership protection across diverse deployment environments .

\noindent{\textbf{Statistical Verifiability.}} 
Experimental results validate the superior ownership verification capability and rigorous statistical reliability of FSW. Across all evaluated backbones, our method consistently yields detection scores $S$ that significantly exceed the theoretical critical thresholds $\tau$. As detailed in Table~\ref{tab:2}, we evaluate the model under various significance levels $\alpha$. For DeepSeek-7B, the margin $S - \tau$ remains as high as $3.0587$ even at an extremely stringent level of $\alpha = 10^{-8}$, effectively rejecting the null hypothesis $H_0$ with near-zero probability of false positives. While Mistral-7B shows a marginal failure at the extreme $\alpha = 10^{-8}$ level, it maintains a robust positive margin of $0.3776$ at $\alpha = 10^{-6}$, confirming high detection confidence. This demonstrates that the functional subspace constructed via GEVP identifies carrier directions with an exceptionally high signal-to-noise ratio (SNR), enabling irrefutable ownership proof that is statistically distinguishable from random fluctuations.

\begin{figure}[t]
\centering
\includegraphics[width=\linewidth]{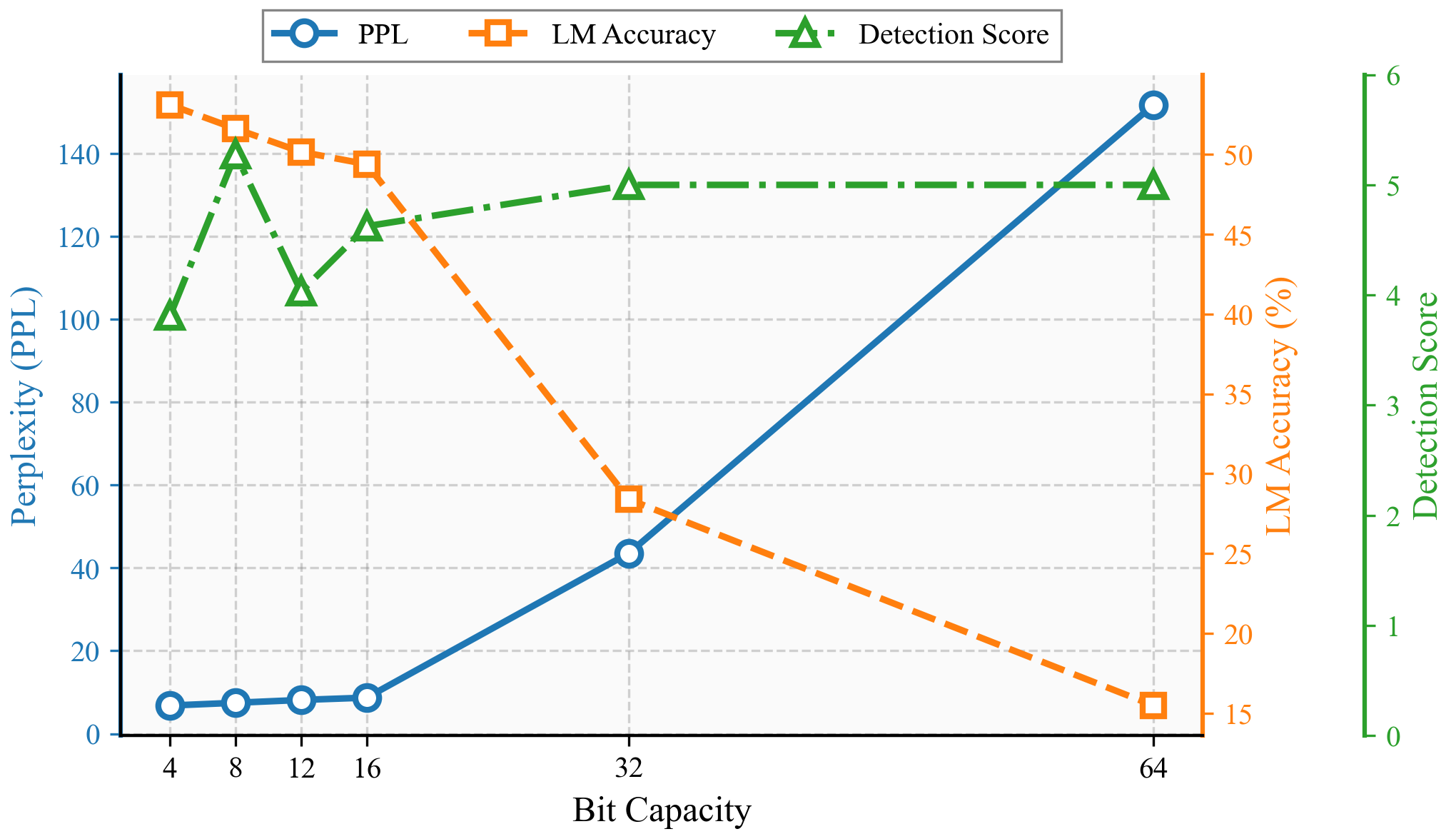}
\caption{
\textbf{Bit Capacity vs. Utility and Detectability}
}
\label{fig:hidden_drift_layerwise}
\end{figure}

\subsection{Robustness Analysis}
\noindent{\textbf{Robustness to Parameter Perturbations.}} 
The resilience of FSW against post-hoc modifications is demonstrated by its performance under parameter-level attacks such as LoRA fine-tuning, quantization, and pruning. As shown in Table \ref{tab:robustness}, the framework maintains, and in some cases even enhances, its detection capabilities under these perturbations; for instance, the detection score for LLaMA-2-7B-hf increased from a baseline of $6.09$ to $6.56$ after INT4 quantization and $6.72$ after magnitude pruning. Furthermore, the watermark proved resistant to LoRA fine-tuning on Meta-LLaMA-3-8B, maintaining an $87.5\%$ bit accuracy while slightly increasing the detection score from $4.00$ to $4.22$. This stability, further supported by the distinct separation of detection scores in Figure \ref{fig:3}, confirms that FSW successfully anchors ownership signals into a functional backbone that remains invariant to common precision losses and structural modifications.

\noindent{\textbf{Resistance to Model Distillation.}} FSW exhibits superior robustness against model distillation attacks compared to current state-of-the-art methods by explicitly decoupling task-critical sensitivity from compression-induced variance. According to Table \ref{tab:distill_baseline_1}, while baseline methods like EmMark often suffer from significant utility degradation, FSW preserves the original model performance, evidenced by the stable PPL of $5.91$ for LLaMA-2-7B after distillation. The "Backbone Distillation" results in Table \ref{tab:robustness} further highlight this efficacy, where DeepSeek-LLM-7B-Chat achieved a perfect $100\%$ bit accuracy even after its weights were re-parameterized. This statistical reliability is underscored in Table \ref{tab:distill_baseline_1} in Appendix \ref{appendix c}, where the detection margin for DeepSeek-7B remained as high as $3.0587$ at an extreme significance level of $\alpha=10^{-8}$, providing irrefutable proof of ownership even under high-compression scenarios.

\subsection{Ablation Studies.}

\begin{table*}[t]
\centering
\small
\setlength{\tabcolsep}{6pt} 
\begin{tabular*}{\textwidth}{@{\extracolsep{\fill}} l ccc ccc ccc}
\toprule
\multirow{2}{*}{\textbf{Variant}} & \multicolumn{3}{c}{\textbf{Detection Score} $S \uparrow$} & \multicolumn{3}{c}{\textbf{Bit Accuracy} $\uparrow$} & \multicolumn{3}{c}{\textbf{Utility (PPL)} $\downarrow$} \\
\cmidrule(lr){2-4} \cmidrule(lr){5-7} \cmidrule(lr){8-10}
& Pre & Post & $\Delta S$ & Pre & Post & $\Delta$Acc & Pre & Post & $\Delta$PPL \\
\midrule
\textbf{Full FSW} & \textbf{6.09} & \textbf{5.84} & \textbf{-0.25} & \textbf{100.0\%} & \textbf{87.5\%} & \textbf{-12.5\%} & \textbf{5.91} & \textbf{5.91} & \textbf{0.00} \\
\midrule
w/o Comp. Inv. & 4.47 & 0.96 & -3.51 & 100.0\% & 100.0\% & 0.00\% & 7.02 & 6.43 & +0.60 \\
w/o Anchored Proj. & 3.08 & -1.94 & -5.02 & 75.0\% & 37.5\% & -37.5\% & 7.31 & 6.82 & +0.49 \\
w/o Consistency Loss & 4.59 & -0.36 & -4.95 & 100.0\% & 37.5\% & -62.5\% & 7.19 & 6.59 & +0.60 \\
w/o Adaptive Thres. & 4.91 & -2.73 & -7.64 & 87.5\% & 12.5\% & -75.0\% & 7.16 & 6.85 & +0.30 \\
\bottomrule
\end{tabular*}
\caption{Ablation study of FSW components on Llama-2-7B-hf. \textit{Pre} and \textit{post} denote metrics before and after the robustness attack. Each variant systematically removes a key design element (Compression Invariance, Anchored Projection, Consistency Loss, or Adaptive Thresholding) to evaluate its individual contribution to robustness and utility.}
\label{tab:ablation}
\end{table*}

\begin{figure}[t]
\centering
\includegraphics[width=\linewidth]{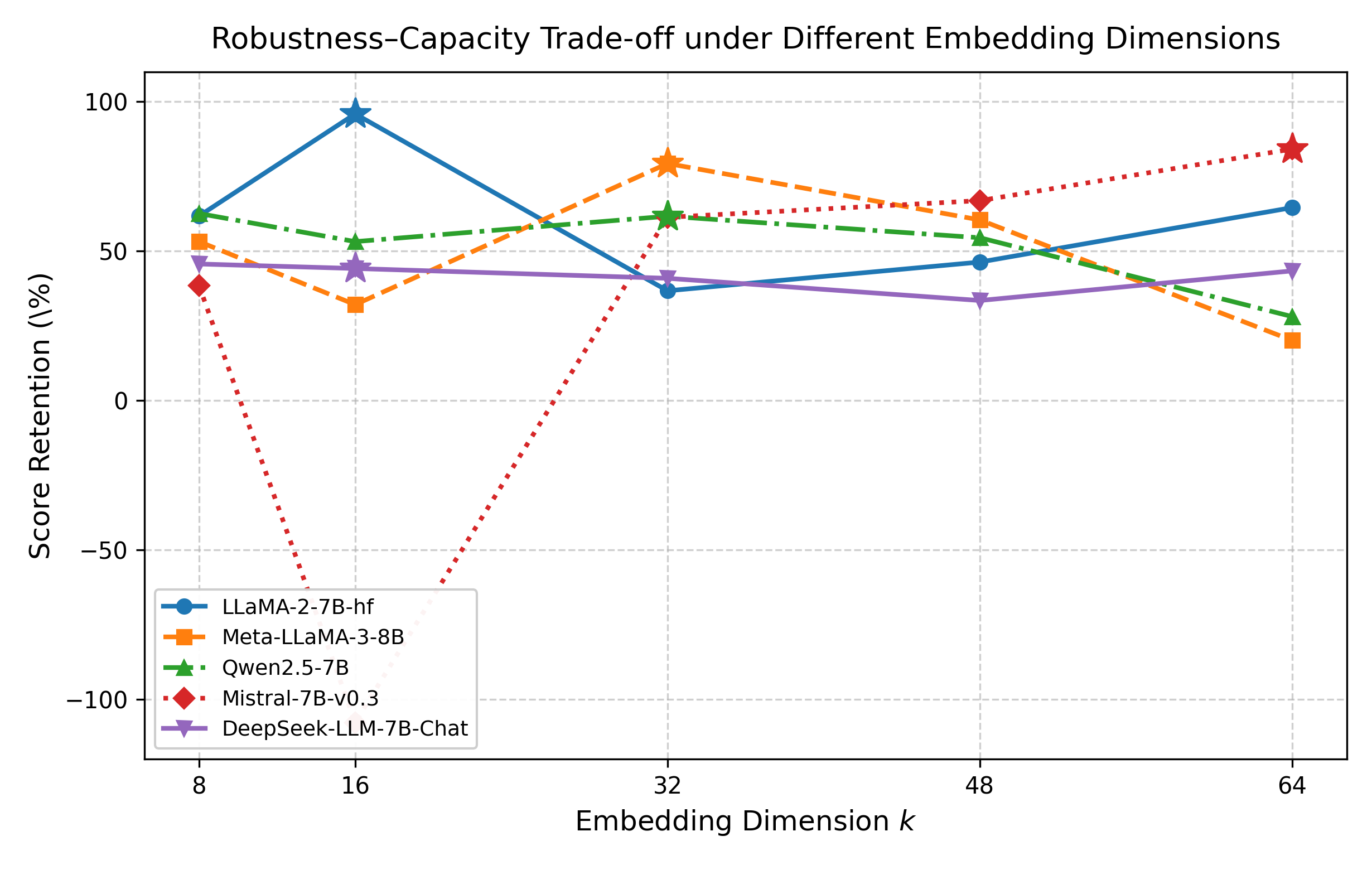}
\caption{
\textbf{Robustness–Capacity Trade-off under Different Embedding Dimensions}
}
\label{fig:hidden_drift_layerwise}
\end{figure}

\noindent{\textbf{Bit Capacity}}
To determine the optimal payload for ownership verification, we analyze the trade-off between watermark capacity, model utility, and detection reliability. As illustrated in Figure \ref{fig:hidden_drift_layerwise}, FSW maintains a stable equilibrium when the bit capacity $B \le 16$, where the PPL remains nearly constant and the LM Accuracy stays at a high level (above $45\%$). However, increasing the capacity to $32$ or $64$ bits leads to a limitation, where the PPL surges to over $150$ and accuracy drops sharply, indicating a structural saturation of the functional subspace. This capacity threshold is further validated by the generalization results in Table \ref{tab:ecc_analysis}, where a configuration within this robust range (typically 16 bits) achieves a perfect Bit Accuracy of 100\% and an AUC of 1.000 on the C4 dataset, despite the increased complexity of the corpus. Notably, while excessive embedding distorts model weights, the Detection Score remains consistently high across all evaluated capacities, confirming that the underlying signal remains recoverable even under high-load conditions. Consequently, 16 bits is identified as the optimal capacity, as it maximizes the information payload for ECC without compromising the model's original functional integrity.

\noindent{\textbf{Ablation Study of Components.}}
The ablation study conducted on the Llama-2-7B-hf model demonstrates that the integration of all FSW components is essential for maintaining the optimal balance between watermark robustness and model utility. As detailed in Table \ref{tab:ablation}, the Full FSW configuration consistently outperforms all variants, maintaining a stable PPL of 5.91 ($\Delta$PPL = 0.00) and a robust post-attack bit accuracy of 87.5\%. In contrast, the removal of the Adaptive Thresholding (w/o Adaptive Thres.) results in the most severe performance degradation, with the post-attack bit accuracy plummeting to 12.5\% and the detection score S dropping to a negative value of -2.73. These results indicate that while Compression Invariance and Consistency Loss are vital for signal persistence, the Anchored Projection and Adaptive Thresholding serve as the foundational pillars for ensuring a stable recovery coordinate system and protecting task-critical representation directions from watermark-induced distortion.

\noindent{\textbf{Ablation Study of Embedding Dimension $k$.}}
The ablation analysis of the embedding dimension $k$ reveals that selecting an optimal subspace size is crucial for maximizing watermark retention without sacrificing model performance. As illustrated in Table \ref{tab:best_k} in Appendix \ref{appendix c}, Table \ref{tab:comprehensive-k-scan} (in Appendix \ref{appendix c}), and Figure \ref{fig:hidden_drift_layerwise}, the optimal $k$ varies significantly across architectures to balance watermark capacity and structural stability, with Llama-2-7B reaching its peak score retention of 95.90\% at $k=16$, while larger models like Mistral-7B require a higher dimension of $k=64$ to achieve a maximum retention rate of 84.15\%. Data from the comprehensive scan in Table 9 indicates that deviations from these optimal values lead to a significant performance trade-off: smaller dimensions often fail to capture sufficient carrier directions, resulting in lower detection scores after robustness attacks, whereas excessively large dimensions begin to overlap with highly sensitive representation regions, causing an unnecessary increase in perplexity. These findings suggest that the optimal $k$ serves as a structural bottleneck that effectively encapsulates the stable functional backbone of each specific LLM, ensuring that the embedded ownership signals are both robust to parameter-level perturbations and non-intrusive to the model's original utility.

\subsection{Cost Analysis}
\noindent{\textbf{Training Overhead Analysis.}}
The evaluation of training overhead reveals that while FSW introduces additional computational requirements compared to standard fine-tuning, the overall costs remain within a manageable range for 7B-scale models. As detailed in Table \ref{tab:overhead}, the total training time for the FSW pipeline on Qwen2.5-7B is 1855.68 seconds across 8001 steps, representing a reasonable increase over the 1101.55 seconds required for standard fine-tuning. Most of this overhead is concentrated in the Phase 4: Watermark Embedding stage, which takes 1764.38 seconds and requires significantly higher GPU memory (56410.34 MB) compared to the baseline (29820.63 MB). This increase in resource consumption is primarily attributed to the dual-model consistency constraints employed during the embedding phase, which necessitate maintaining additional representation states to ensure that the watermarking process does not degrade the original model's functional utility.

\section{Conclusion}
This paper presents the \texttt{\textbf{FSW}}, a framework for LLMs ownership protection. To address watermark failure caused by fine-tuning, quantization, and distillation, we construct a low-dimensional functional subspace that combines task-criticality with compression-invariance by solving the GEVP. This method embeds the watermark into the model's core representation directions to ensure robustness under parameter-level modifications. To balance robustness and model utility, we introduce an adaptive spectral truncation strategy to avoid performance degradation and utilize vector consistency constraints to preserve the original semantic distribution. Extensive experiments have shown that \texttt{\textbf{FSW}} effectively defends against various model attacks while maintaining an extremely low false positive rate. This provides a rigorous technical approach for the intellectual property protection of large-scale models.

\section*{Impact Statement}
This work aims to advance research on model-side watermarking for large language models by improving robustness under common model modifications. The proposed method is intended to support model ownership verification and provenance analysis. We do not anticipate significant negative societal impacts arising from this work.


\bibliography{example_paper}

@inproceedings{kirchenbauer2023watermark,
  title={A watermark for large language models},
  author={Kirchenbauer, John and Geiping, Jonas and Wen, Yuxin and Katz, Jonathan and Miers, Ian and Goldstein, Tom},
  booktitle={International Conference on Machine Learning},
  pages={17061--17084},
  year={2023},
  organization={PMLR}
}

@article{dathathri2024scalable,
  title={Scalable watermarking for identifying large language model outputs},
  author={Dathathri, Sumanth and See, Abigail and Ghaisas, Sumedh and Huang, Po-Sen and McAdam, Rob and Welbl, Johannes and Bachani, Vandana and Kaskasoli, Alex and Stanforth, Robert and Matejovicova, Tatiana and others},
  journal={Nature},
  volume={634},
  number={8035},
  pages={818--823},
  year={2024},
  publisher={Nature Publishing Group UK London}
}

@inproceedings{zhang2024remark,
  title={$\{$REMARK-LLM$\}$: A robust and efficient watermarking framework for generative large language models},
  author={Zhang, Ruisi and Hussain, Shehzeen Samarah and Neekhara, Paarth and Koushanfar, Farinaz},
  booktitle={33rd USENIX Security Symposium (USENIX Security 24)},
  pages={1813--1830},
  year={2024}
}

@inproceedings{dabiriaghdam-wang-2025-simmark,
    title = "{S}im{M}ark: A Robust Sentence-Level Similarity-Based Watermarking Algorithm for Large Language Models",
    author = "Dabiriaghdam, Amirhossein  and
      Wang, Lele",
    editor = "Christodoulopoulos, Christos  and
      Chakraborty, Tanmoy  and
      Rose, Carolyn  and
      Peng, Violet",
    booktitle = "Proceedings of the 2025 Conference on Empirical Methods in Natural Language Processing",
    month = nov,
    year = "2025",
    address = "Suzhou, China",
    publisher = "Association for Computational Linguistics",
    url = "https://aclanthology.org/2025.emnlp-main.1567/",
    doi = "10.18653/v1/2025.emnlp-main.1567",
    pages = "30785--30806",
    ISBN = "979-8-89176-332-6"
}

@inproceedings{mao2025watermarking,
  title={Watermarking large language models: An unbiased and low-risk method},
  author={Mao, Minjia and Wei, Dongjun and Chen, Zeyu and Fang, Xiao and Chau, Michael},
  booktitle={Proceedings of the 63rd Annual Meeting of the Association for Computational Linguistics (Volume 1: Long Papers)},
  pages={7939--7960},
  year={2025}
}

@inproceedings{xurobust,
  title={Robust Multi-bit Text Watermark with LLM-based Paraphrasers},
  author={Xu, Xiaojun and Jia, Jinghan and Yao, Yuanshun and Liu, Yang and Li, Hang},
  booktitle={Forty-second International Conference on Machine Learning},
  year={2025}
}

@inproceedings{niess2025ensemble,
  title={Ensemble watermarks for large language models},
  author={Niess, Georg and Kern, Roman},
  booktitle={Proceedings of the 63rd Annual Meeting of the Association for Computational Linguistics (Volume 1: Long Papers)},
  pages={2903--2916},
  year={2025}
}

@inproceedings{liu2025context,
  title={In-Context Watermarks for Large Language Models},
  author={Liu, Yepeng and Zhao, Xuandong and Kruegel, Christopher and Song, Dawn and Bu, Yuheng},
  booktitle={ICML 2025 Workshop on Reliable and Responsible Foundation Models},
  year={2025}
}

@inproceedings{hetheoretically,
  title={Theoretically Grounded Framework for LLM Watermarking: A Distribution-Adaptive Approach},
  author={He, Haiyun and Liu, Yepeng and Wang, Ziqiao and Mao, Yongyi and Bu, Yuheng},
  booktitle={The 1st Workshop on GenAI Watermarking},
  year={2024}
}

@article{sander2024watermarking,
  title={Watermarking makes language models radioactive},
  author={Sander, Tom and Fernandez, Pierre and Durmus, Alain and Douze, Matthijs and Furon, Teddy},
  journal={Advances in Neural Information Processing Systems},
  volume={37},
  pages={21079--21113},
  year={2024}
}

@inproceedings{li2023watermarking,
  title={Watermarking LLMs with Weight Quantization},
  author={Li, Linyang and Jiang, Botian and Wang, Pengyu and Ren, Ke and Yan, Hang and Qiu, Xipeng},
  booktitle={Findings of the Association for Computational Linguistics: EMNLP 2023},
  pages={3368--3378},
  year={2023}
}

@article{dai2025seal,
  title={SEAL: Subspace-Anchored Watermarks for LLM Ownership},
  author={Dai, Yanbo and Li, Zongjie and Ji, Zhenlan and Wang, Shuai},
  journal={arXiv preprint arXiv:2511.11356},
  year={2025}
}

@inproceedings{yamabe2025mergeprint,
  title={MERGEPRINT: Merge-Resistant Fingerprints for Robust Black-box Ownership Verification of Large Language Models},
  author={Yamabe, Shojiro and Waseda, Futa Kai and Takahashi, Tsubasa and Wataoka, Koki},
  booktitle={Proceedings of the 63rd Annual Meeting of the Association for Computational Linguistics (Volume 1: Long Papers)},
  pages={6894--6916},
  year={2025}
}

@inproceedings{darvish2019deepsigns,
  title={Deepsigns: An end-to-end watermarking framework for ownership protection of deep neural networks},
  author={Darvish Rouhani, Bita and Chen, Huili and Koushanfar, Farinaz},
  booktitle={Proceedings of the twenty-fourth international conference on architectural support for programming languages and operating systems},
  pages={485--497},
  year={2019}
}

@inproceedings{adi2018turning,
  title={Turning your weakness into a strength: Watermarking deep neural networks by backdooring},
  author={Adi, Yossi and Baum, Carsten and Cisse, Moustapha and Pinkas, Benny and Keshet, Joseph},
  booktitle={27th USENIX security symposium (USENIX Security 18)},
  pages={1615--1631},
  year={2018}
}

@misc{meta2025llama4_multimodal_intelligence,
  title        = {The Llama 4 herd: The beginning of a new era of natively multimodal AI innovation},
  author       = {{Meta AI}},
  howpublished = {\url{https://ai.meta.com/blog/llama-4-multimodal-intelligence/}},
  year         = {2025},
}

@misc{mistral2025mistral3,
  title        = {Introducing Mistral 3},
  author       = {{Mistral AI}},
  year         = {2025},
  month        = dec,
  day          = {2},
  howpublished = {\url{https://mistral.ai/news/mistral-3}},
}

@misc{yang2025qwen3technicalreport,
      title={Qwen3 Technical Report}, 
      author={An Yang and Anfeng Li and Baosong Yang and Beichen Zhang and Binyuan Hui and Bo Zheng and Bowen Yu and Chang Gao and Chengen Huang and Chenxu Lv and Chujie Zheng and Dayiheng Liu and Fan Zhou and Fei Huang and Feng Hu and Hao Ge and Haoran Wei and Huan Lin and Jialong Tang and Jian Yang and Jianhong Tu and Jianwei Zhang and Jianxin Yang and Jiaxi Yang and Jing Zhou and Jingren Zhou and Junyang Lin and Kai Dang and Keqin Bao and Kexin Yang and Le Yu and Lianghao Deng and Mei Li and Mingfeng Xue and Mingze Li and Pei Zhang and Peng Wang and Qin Zhu and Rui Men and Ruize Gao and Shixuan Liu and Shuang Luo and Tianhao Li and Tianyi Tang and Wenbiao Yin and Xingzhang Ren and Xinyu Wang and Xinyu Zhang and Xuancheng Ren and Yang Fan and Yang Su and Yichang Zhang and Yinger Zhang and Yu Wan and Yuqiong Liu and Zekun Wang and Zeyu Cui and Zhenru Zhang and Zhipeng Zhou and Zihan Qiu},
      year={2025},
      eprint={2505.09388},
      archivePrefix={arXiv},
      primaryClass={cs.CL},
      url={https://arxiv.org/abs/2505.09388}, 
}

@misc{gemmateam2025gemma3technicalreport,
      title={Gemma 3 Technical Report}, 
      author={Gemma Team and Aishwarya Kamath and Johan Ferret and Shreya Pathak and Nino Vieillard and Ramona Merhej and Sarah Perrin and Tatiana Matejovicova and Alexandre Ramé and Morgane Rivière and Louis Rouillard and Thomas Mesnard and Geoffrey Cideron and Jean-bastien Grill and Sabela Ramos and Edouard Yvinec and Michelle Casbon and Etienne Pot and Ivo Penchev and Gaël Liu and Francesco Visin and Kathleen Kenealy and Lucas Beyer and Xiaohai Zhai and Anton Tsitsulin and Robert Busa-Fekete and Alex Feng and Noveen Sachdeva and Benjamin Coleman and Yi Gao and Basil Mustafa and Iain Barr and Emilio Parisotto and David Tian and Matan Eyal and Colin Cherry and Jan-Thorsten Peter and Danila Sinopalnikov and Surya Bhupatiraju and Rishabh Agarwal and Mehran Kazemi and Dan Malkin and Ravin Kumar and David Vilar and Idan Brusilovsky and Jiaming Luo and Andreas Steiner and Abe Friesen and Abhanshu Sharma and Abheesht Sharma and Adi Mayrav Gilady and Adrian Goedeckemeyer and Alaa Saade and Alex Feng and Alexander Kolesnikov and Alexei Bendebury and Alvin Abdagic and Amit Vadi and András György and André Susano Pinto and Anil Das and Ankur Bapna and Antoine Miech and Antoine Yang and Antonia Paterson and Ashish Shenoy and Ayan Chakrabarti and Bilal Piot and Bo Wu and Bobak Shahriari and Bryce Petrini and Charlie Chen and Charline Le Lan and Christopher A. Choquette-Choo and CJ Carey and Cormac Brick and Daniel Deutsch and Danielle Eisenbud and Dee Cattle and Derek Cheng and Dimitris Paparas and Divyashree Shivakumar Sreepathihalli and Doug Reid and Dustin Tran and Dustin Zelle and Eric Noland and Erwin Huizenga and Eugene Kharitonov and Frederick Liu and Gagik Amirkhanyan and Glenn Cameron and Hadi Hashemi and Hanna Klimczak-Plucińska and Harman Singh and Harsh Mehta and Harshal Tushar Lehri and Hussein Hazimeh and Ian Ballantyne and Idan Szpektor and Ivan Nardini and Jean Pouget-Abadie and Jetha Chan and Joe Stanton and John Wieting and Jonathan Lai and Jordi Orbay and Joseph Fernandez and Josh Newlan and Ju-yeong Ji and Jyotinder Singh and Kat Black and Kathy Yu and Kevin Hui and Kiran Vodrahalli and Klaus Greff and Linhai Qiu and Marcella Valentine and Marina Coelho and Marvin Ritter and Matt Hoffman and Matthew Watson and Mayank Chaturvedi and Michael Moynihan and Min Ma and Nabila Babar and Natasha Noy and Nathan Byrd and Nick Roy and Nikola Momchev and Nilay Chauhan and Noveen Sachdeva and Oskar Bunyan and Pankil Botarda and Paul Caron and Paul Kishan Rubenstein and Phil Culliton and Philipp Schmid and Pier Giuseppe Sessa and Pingmei Xu and Piotr Stanczyk and Pouya Tafti and Rakesh Shivanna and Renjie Wu and Renke Pan and Reza Rokni and Rob Willoughby and Rohith Vallu and Ryan Mullins and Sammy Jerome and Sara Smoot and Sertan Girgin and Shariq Iqbal and Shashir Reddy and Shruti Sheth and Siim Põder and Sijal Bhatnagar and Sindhu Raghuram Panyam and Sivan Eiger and Susan Zhang and Tianqi Liu and Trevor Yacovone and Tyler Liechty and Uday Kalra and Utku Evci and Vedant Misra and Vincent Roseberry and Vlad Feinberg and Vlad Kolesnikov and Woohyun Han and Woosuk Kwon and Xi Chen and Yinlam Chow and Yuvein Zhu and Zichuan Wei and Zoltan Egyed and Victor Cotruta and Minh Giang and Phoebe Kirk and Anand Rao and Kat Black and Nabila Babar and Jessica Lo and Erica Moreira and Luiz Gustavo Martins and Omar Sanseviero and Lucas Gonzalez and Zach Gleicher and Tris Warkentin and Vahab Mirrokni and Evan Senter and Eli Collins and Joelle Barral and Zoubin Ghahramani and Raia Hadsell and Yossi Matias and D. Sculley and Slav Petrov and Noah Fiedel and Noam Shazeer and Oriol Vinyals and Jeff Dean and Demis Hassabis and Koray Kavukcuoglu and Clement Farabet and Elena Buchatskaya and Jean-Baptiste Alayrac and Rohan Anil and Dmitry and Lepikhin and Sebastian Borgeaud and Olivier Bachem and Armand Joulin and Alek Andreev and Cassidy Hardin and Robert Dadashi and Léonard Hussenot},
      year={2025},
      eprint={2503.19786},
      archivePrefix={arXiv},
      primaryClass={cs.CL},
      url={https://arxiv.org/abs/2503.19786}, 
}

@article{shao2025sok,
  title={Sok: Large language model copyright auditing via fingerprinting},
  author={Shao, Shuo and Li, Yiming and He, Yu and Yao, Hongwei and Yang, Wenyuan and Tao, Dacheng and Qin, Zhan},
  journal={arXiv preprint arXiv:2508.19843},
  year={2025}
}

@inproceedings{pasquini2025llmmap,
  title={$\{$LLMmap$\}$: Fingerprinting for large language models},
  author={Pasquini, Dario and Kornaropoulos, Evgenios M and Ateniese, Giuseppe},
  booktitle={34th USENIX Security Symposium (USENIX Security 25)},
  pages={299--318},
  year={2025}
}

@inproceedings{wang2021riga,
  title={Riga: Covert and robust white-box watermarking of deep neural networks},
  author={Wang, Tianhao and Kerschbaum, Florian},
  booktitle={Proceedings of the web conference 2021},
  pages={993--1004},
  year={2021}
}

@inproceedings{xu2024instructional,
  title={Instructional fingerprinting of large language models},
  author={Xu, Jiashu and Wang, Fei and Ma, Mingyu and Koh, Pang Wei and Xiao, Chaowei and Chen, Muhao},
  booktitle={Proceedings of the 2024 Conference of the North American Chapter of the Association for Computational Linguistics: Human Language Technologies (Volume 1: Long Papers)},
  pages={3277--3306},
  year={2024}
}

@inproceedings{yao2024promptcare,
  title={Promptcare: Prompt copyright protection by watermark injection and verification},
  author={Yao, Hongwei and Lou, Jian and Qin, Zhan and Ren, Kui},
  booktitle={2024 IEEE Symposium on Security and Privacy (SP)},
  pages={845--861},
  year={2024},
  organization={IEEE}
}

@inproceedings{yan2023rethinking,
  title={Rethinking $\{$White-Box$\}$ watermarks on deep learning models under neural structural obfuscation},
  author={Yan, Yifan and Pan, Xudong and Zhang, Mi and Yang, Min},
  booktitle={32nd USENIX Security Symposium (USENIX Security 23)},
  pages={2347--2364},
  year={2023}
}

@inproceedings{li2025differentiation,
  title={Differentiation-based extraction of proprietary data from fine-tuned llms},
  author={Li, Zongjie and Wu, Daoyuan and Wang, Shuai and Su, Zhendong},
  booktitle={Proceedings of the 2025 ACM SIGSAC Conference on Computer and Communications Security},
  pages={3071--3085},
  year={2025}
}

@article{touvron2023llama,
  title={Llama 2: Open foundation and fine-tuned chat models},
  author={Touvron, Hugo and Martin, Louis and Stone, Kevin and Albert, Peter and Almahairi, Amjad and Babaei, Yasmine and Bashlykov, Nikolay and Batra, Soumya and Bhargava, Prajjwal and Bhosale, Shruti and others},
  journal={arXiv preprint arXiv:2307.09288},
  year={2023}
}

@article{roque2024evolution,
  title={The Evolution of Llama: From Llama 1 to Llama 3.1},
  author={Roque, Lu{\'\i}s},
  journal={Medium},
  year={2024}
}

@article{team2024qwen2,
  title={Qwen2 technical report},
  author={Team, Qwen and others},
  journal={arXiv preprint arXiv:2407.10671},
  volume={2},
  number={3},
  year={2024}
}

@misc{jiang2023mistral7b,
      title={Mistral 7B}, 
      author={Albert Q. Jiang and Alexandre Sablayrolles and Arthur Mensch and Chris Bamford and Devendra Singh Chaplot and Diego de las Casas and Florian Bressand and Gianna Lengyel and Guillaume Lample and Lucile Saulnier and Lélio Renard Lavaud and Marie-Anne Lachaux and Pierre Stock and Teven Le Scao and Thibaut Lavril and Thomas Wang and Timothée Lacroix and William El Sayed},
      year={2023},
      eprint={2310.06825},
      archivePrefix={arXiv},
      primaryClass={cs.CL},
      url={https://arxiv.org/abs/2310.06825}, 
}

@article{bi2024deepseek,
  title={Deepseek llm: Scaling open-source language models with longtermism},
  author={Bi, Xiao and Chen, Deli and Chen, Guanting and Chen, Shanhuang and Dai, Damai and Deng, Chengqi and Ding, Honghui and Dong, Kai and Du, Qiushi and Fu, Zhe and others},
  journal={arXiv preprint arXiv:2401.02954},
  year={2024}
}

@article{zhu2023multimodal,
  title={Multimodal c4: An open, billion-scale corpus of images interleaved with text},
  author={Zhu, Wanrong and Hessel, Jack and Awadalla, Anas and Gadre, Samir Yitzhak and Dodge, Jesse and Fang, Alex and Yu, Youngjae and Schmidt, Ludwig and Wang, William Yang and Choi, Yejin},
  journal={Advances in Neural Information Processing Systems},
  volume={36},
  pages={8958--8974},
  year={2023}
}

@article{zellers2019hellaswag,
  title={Hellaswag: Can a machine really finish your sentence?},
  author={Zellers, Rowan and Holtzman, Ari and Bisk, Yonatan and Farhadi, Ali and Choi, Yejin},
  journal={arXiv preprint arXiv:1905.07830},
  year={2019}
}

@article{clark2018think,
  title={Think you have solved question answering? try arc, the ai2 reasoning challenge},
  author={Clark, Peter and Cowhey, Isaac and Etzioni, Oren and Khot, Tushar and Sabharwal, Ashish and Schoenick, Carissa and Tafjord, Oyvind},
  journal={arXiv preprint arXiv:1803.05457},
  year={2018}
}

@inproceedings{liu2025evolution,
  title={Evolution of the Spectral Dimension of Transformer Activations},
  author={Liu, Andy Zeyi and Paquette, Elliot and Sous, John},
  booktitle={OPT 2025: Optimization for Machine Learning}
}

@inproceedings{zhang2024emmark,
  title={EmMark: Robust watermarks for IP protection of embedded quantized large language models},
  author={Zhang, Ruisi and Koushanfar, Farinaz},
  booktitle={Proceedings of the 61st ACM/IEEE Design Automation Conference},
  pages={1--6},
  year={2024}
}
\bibliographystyle{icml2026}

\newpage
\appendix
\onecolumn
\section{Algorithm.}
\label{appendix a}
\begin{algorithm}[ht]
\caption{Functional Subspace Watermarking (\texttt{FSW}) Implementation Pipeline}
\label{alg:fsw_full}
\begin{algorithmic}[1]
\STATE \textbf{Input:} Pre-trained model $f_{\theta_0}$, calibration set $\mathcal{D}$, challenge set $\mathcal{C}$, layer $\ell$, subspace dimension $k$, margin $\gamma$, target bits $\{y_j\}_{j=1}^M$.
\STATE \textbf{Phase 1: Representation Analysis}
\STATE \quad Extract hidden states $r(x; \theta_0)$ from layer $\ell$ for $x \sim \mathcal{D}$ and compute mean $\mu_{\mathcal{D}}$.
\STATE \quad Estimate Fisher matrix $F$ (Eq. 2) and Invariance matrix $C_{\mathcal{A}}$ (Eq. 3) via stochastic operators $\mathcal{A}$.
\STATE \textbf{Phase 2: Subspace Construction \& Optimization}
\STATE \quad Solve Generalized Eigenvalue Problem (GEVP): $F u_i = \lambda_i C_{\mathcal{A}} u_i$ to obtain eigenvectors $\{u_i\}$.
\STATE \quad Determine spectral sweet spot $\mathcal{I}$ via adaptive truncation thresholds $\tau_{\mathrm{lower}}, \tau_{\mathrm{upper}}$ (Eq. 5).
\STATE \quad Construct the functional backbone projection matrix $U^\star = [u_{i_1}, \dots, u_{i_k}]$ where $i \in \mathcal{I}$.
\STATE \textbf{Phase 3: Constrained Watermark Embedding}
\STATE \quad Generate $M$ mutually orthogonal secret key vectors $\{b_1, \dots, b_M\} \in \mathbb{R}^k$.
\STATE \quad Fine-tune model $f_\theta$ by minimizing the joint objective (Eq. 10):
\STATE \quad \quad $L = L_{\mathrm{LM}} + \lambda_{\mathrm{wm}} L_{\mathrm{wm}} + \lambda_{\mathrm{con}} L_{\mathrm{con}}$.
\STATE \textbf{Phase 4: Statistical Ownership Verification}
\STATE \quad Given a suspected model, project representations into $U^\star$ and compute $S$ (Eq. 11).
\STATE \quad Perform hypothesis test: reject $H_0$ if $\text{FPR} = \frac{1}{2} \text{erfc}(\frac{T}{\sqrt{2}\sigma_0}) < \alpha$ (threshold $\alpha$).
\STATE \textbf{Phase 5: Multi-bit Message Decoding}
\STATE \quad Compute per-bit projection statistics $S_j = \frac{1}{|\mathcal{C}|} \sum_{x \in \mathcal{C}} \frac{b_j^\top z(x)}{\|b_j\|_2}$ for $j=1,\dots,M$.
\STATE \quad \textbf{Return:} Extracted bits $\hat y_j = \text{sign}(S_j)$ and final ECC-decoded message.
\label{algorithm}
\end{algorithmic}
\end{algorithm}

\section{Detailed Experimental Setup}
\label{app:exp_details}

\subsection{Model Selection and Specifications}
The model selection is shown in Table ~\ref{tab:model_specs} in Appendix \ref{appendix c}.



\subsection{Dataset Descriptions and Allocation}
We combine multiple datasets for subspace estimation, fine-tuning, and functional evaluation to ensure that this is not a coincidence.

\begin{itemize}
    \item \textbf{WikiText-2:} A high-quality collection of Wikipedia articles. We use 500 samples for estimating the Fisher matrix $F$ and invariance matrix $C_{\mathcal{A}}$, and 500 separate samples for evaluating language modeling perplexity (PPL).
    \item \textbf{C4 (Colossal Clean Crawled Corpus):} A massive, cleaned version of Common Crawl. We sample 3,000 sequences for watermark embedding via LoRA fine-tuning to enhance distributional robustness.
    \item \textbf{HellaSwag \& ARC-Easy:} Used as downstream benchmarks (500 samples each) to assess common-sense reasoning and knowledge retention after watermarking.
\end{itemize}

\subsection{Implementation Details}
\label{Implementation Details}
The proposed framework is implemented using PyTorch and the PEFT library. We extract the last-token hidden representations from the middle layer of each model as specified in Table~\ref{tab:model_specs}. For the invariance matrix $C_{\mathcal{A}}$, we apply $N=3$ stochastic perturbations per sample using a family of operators $\mathcal{A}$:
\begin{itemize}
    \item \textbf{Linear Projection:} rank ratio of 0.25 relative to the hidden dimension.
    \item \textbf{Quantization Noise:} Gaussian noise with $\sigma=0.1$.
    \item \textbf{Structural Dropout:} Bernoulli mask with retention rate $p=0.9$.
\end{itemize}

The functional backbone $U^\star$ is constructed with $k=32$, using adaptive thresholds $\tau_{\mathrm{lower}}=10^{-4}$ and $\tau_{\mathrm{upper}}=0.6$. LoRA fine-tuning uses rank $r=16$ and $\alpha=32$. The joint objective is optimized with $\lambda_{\mathrm{wm}}=10$, $\lambda_{\mathrm{con}}=0.1$, and a margin $\gamma=5.0$. All experiments are conducted on a single NVIDIA RTX 6000 Ada GPU.

\section{Supplementary Table}

\begin{table*}[t]
\centering
\small
\setlength{\tabcolsep}{8pt} 
\begin{tabular*}{\textwidth}{@{\extracolsep{\fill}} ll ccc ccc}
\toprule
\multirow{2.5}{*}{\textbf{Model}} & \multirow{2.5}{*}{\textbf{Variant}} & \multicolumn{3}{c}{\textbf{Original (Pre-Attack)}} & \multicolumn{3}{c}{\textbf{Attacked (Post-Attack)}} \\
\cmidrule(lr){3-5} \cmidrule(lr){6-8}
& & PPL $\downarrow$ & Hella $\uparrow$ & ARC $\uparrow$ & PPL $\downarrow$ & Hella $\uparrow$ & ARC $\uparrow$ \\
\midrule

\multirow{5}{*}{\textbf{Llama-2-7B}} 
& Clean FT (No WM)    & 5.30    & 66.40\% & 71.60\% & ---     & ---     & ---     \\
& Full DIFSW          & 5.91    & 64.80\% & 71.00\% & 5.91    & 65.00\% & 71.20\% \\
& EmMark              & 416.15  & 65.80\% & 71.40\% & 5.73    & 65.80\% & 72.60\% \\
& Weight Quant.       & 20.88   & 66.60\% & 71.60\% & 5.78    & 68.00\% & 72.40\% \\
& Naive Top-k         & 7.30    & 66.80\% & 72.20\% & 6.75    & 66.60\% & 71.60\% \\
\midrule

\multirow{5}{*}{\textbf{Llama-3-8B}} 
& Clean FT (No WM)    & 5.83    & 68.60\% & 79.60\% & ---     & ---     & ---     \\
& Full DIFSW          & 6.20    & 67.80\% & 75.80\% & 6.12    & 68.00\% & 76.00\% \\
& EmMark              & 23.24   & 68.60\% & 77.40\% & 6.09    & 69.00\% & 77.80\% \\
& Weight Quant.       & 9.77    & 69.60\% & 76.00\% & 6.03    & 69.80\% & 77.40\% \\
& Naive Top-k         & 7.35    & 69.80\% & 74.20\% & 6.64    & 70.00\% & 76.40\% \\
\midrule

\multirow{5}{*}{\textbf{Qwen2.5-7B}} 
& Clean FT (No WM)    & 6.35    & 67.00\% & 75.80\% & ---     & ---     & ---     \\
& Full DIFSW          & 7.24    & 66.80\% & 73.40\% & 7.66    & 66.80\% & 73.20\% \\
& EmMark              & 40.62   & 67.00\% & 72.00\% & 6.68    & 67.20\% & 78.60\% \\
& Weight Quant.       & 24.02   & 66.80\% & 78.80\% & 6.66    & 66.80\% & 78.80\% \\
& Naive Top-k         & 7.94    & 67.00\% & 77.20\% & 7.79    & 67.00\% & 77.00\% \\
\midrule

\multirow{5}{*}{\textbf{Mistral-7B}} 
& Clean FT (No WM)    & 16.68   & 68.60\% & 77.00\% & ---     & ---     & ---     \\
& Full DIFSW          & 17.25   & 69.00\% & 77.00\% & 6.59    & 68.60\% & 76.40\% \\
& EmMark              & 29.20   & 71.80\% & 78.40\% & 6.24    & 70.60\% & 77.20\% \\
& Weight Quant.       & 27.49   & 71.40\% & 79.60\% & 6.23    & 69.20\% & 76.60\% \\
& Naive Top-k         & 21.24   & 70.40\% & 76.40\% & 6.93    & 67.20\% & 73.20\% \\
\midrule

\multirow{5}{*}{\textbf{DeepSeek-7B}} 
& Clean FT (No WM)    & 59.52   & 67.40\% & 72.20\% & ---     & ---     & ---     \\
& Full DIFSW          & 77.31   & 60.40\% & 66.20\% & 61.39   & 60.60\% & 66.60\% \\
& EmMark              & 152.36  & 69.60\% & 70.60\% & 7.95    & 70.20\% & 72.40\% \\
& Weight Quant.       & 129.89  & 71.00\% & 70.20\% & 8.23    & 70.60\% & 71.20\% \\
& Naive Top-k         & 87.12   & 67.40\% & 70.40\% & 67.00   & 67.00\% & 70.80\% \\
\bottomrule
\end{tabular*}
\caption{\textbf{Baseline Comparison under Model Distillation.} We report perplexity (PPL) and downstream accuracies (HellaSwag and ARC-Challenge) before and after distillation for multiple watermarking baselines.}
\label{tab:distill_baseline_1}
\end{table*}
\begin{table*}[t]
\centering
\small
\setlength{\tabcolsep}{8pt} 
\begin{tabular*}{\textwidth}{@{\extracolsep{\fill}} ll ccc ccc c}
\toprule
\multirow{2.5}{*}{\textbf{Model}} & \multirow{2.5}{*}{\bm{$k$}} & \multicolumn{3}{c}{\textbf{Original (Pre-Attack)}} & \multicolumn{3}{c}{\textbf{Attacked (Post-Attack)}} & \multirow{2.5}{*}{\textbf{Ret. (\%)}} \\
\cmidrule(lr){3-5} \cmidrule(lr){6-8}
& & PPL $\downarrow$ & Score $\uparrow$ & Acc $\uparrow$ & PPL $\downarrow$ & Score $\uparrow$ & Acc $\uparrow$ & \\
\midrule

\multirow{5}{*}{\textbf{Llama-2-7B}} 
& 8  & 5.9661 & 5.0625 & 100.0\% & 6.0652 & 3.1250 & 50.0\% & 61.73 \\
& \textbf{16} & 5.9133 & 6.0938 & 100.0\% & 5.9139 & 5.8438 & 87.5\% & \textbf{95.90} \\
& 32 & 6.0213 & 4.9375 & 100.0\% & 6.0903 & 1.8125 & 62.5\% & 36.71 \\
& 48 & 5.9709 & 5.9375 & 87.5\%  & 6.0322 & 2.7500 & 62.5\% & 46.32 \\
& 64 & 5.9611 & 5.1563 & 87.5\%  & 6.2732 & 3.3281 & 87.5\% & 64.55 \\
\midrule

\multirow{5}{*}{\textbf{Llama-3-8B}} 
& 8  & 6.1810 & 5.4063 & 100.0\% & 6.0825 & 2.8750 & 87.5\% & 53.18 \\
& 16 & 6.1936 & 5.1875 & 87.5\%  & 6.0938 & 1.6641 & 75.0\% & 32.08 \\
& \textbf{32} & 6.2016 & 4.0000 & 100.0\% & 6.1220 & 3.1719 & 62.5\% & \textbf{79.30} \\
& 48 & 6.1880 & 4.5000 & 87.5\%  & 6.0819 & 2.7188 & 62.5\% & 60.42 \\
& 64 & 6.1874 & 5.2500 & 100.0\% & 6.1162 & 1.0547 & 75.0\% & 20.09 \\
\midrule

\multirow{5}{*}{\textbf{Qwen2.5-7B}} 
& 8  & 7.4219 & 5.5938 & 100.0\% & 7.9128 & 3.5000 & 62.5\% & 62.57 \\
& 16 & 7.1660 & 4.4375 & 100.0\% & 7.6541 & 2.3594 & 87.5\% & 53.17 \\
& \textbf{32} & 7.2437 & 5.0938 & 100.0\% & 7.6610 & 3.1406 & 75.0\% & \textbf{61.66} \\
& 48 & 7.4752 & 4.9063 & 100.0\% & 7.9819 & 2.6719 & 62.5\% & 54.46 \\
& 64 & 7.1649 & 5.3438 & 100.0\% & 7.8791 & 1.5000 & 62.5\% & 28.07 \\
\midrule

\multirow{5}{*}{\textbf{Mistral-7B}} 
& 8  & 17.4032 & 4.1250 & 100.0\% & 6.5614 & 1.5859  & 62.5\% & 38.45 \\
& 16 & 17.2952 & 4.0313 & 100.0\% & 6.5578 & -4.3438 & 25.0\% & -107.75 \\
& 32 & 17.2976 & 5.2500 & 100.0\% & 6.5754 & 3.2188  & 87.5\% & 61.31 \\
& 48 & 17.3229 & 5.0000 & 100.0\% & 6.7003 & 3.3438  & 75.0\% & 66.88 \\
& \textbf{64} & 17.2461 & 3.8438 & 100.0\% & 6.5854 & 3.2344  & 75.0\% & \textbf{84.15} \\
\midrule

\multirow{5}{*}{\textbf{DeepSeek-7B}} 
& 8  & 76.7636 & 3.7969 & 100.0\% & 58.3845 & 1.7344 & 100.0\% & 45.68 \\
& \textbf{16} & 77.3068 & 3.7500 & 100.0\% & 61.3864 & 1.6563 & 100.0\% & \textbf{44.17} \\
& 32 & 75.3895 & 3.8594 & 100.0\% & 68.7011 & 1.5781 & 100.0\% & 40.89 \\
& 48 & 80.5563 & 4.2500 & 100.0\% & 66.9643 & 1.4219 & 100.0\% & 33.46 \\
& 64 & 80.3019 & 4.1250 & 100.0\% & 71.6761 & 1.7891 & 100.0\% & 43.37 \\
\bottomrule
\end{tabular*}
\caption{Comprehensive Scan of Embedding Dimension $k$ across Different LLM Backbones. \textit{Pre} and \textit{post} denote metrics before and after the robustness attack (e.g., distillation or quantization). Ret.(\%) indicates the score retention rate. \textbf{Bold} $k$ values represent the optimal dimensions selected for each model.}
\label{tab:comprehensive-k-scan}
\end{table*}

\begin{table*}[t]
\centering
\small
\setlength{\tabcolsep}{10pt} 
\begin{tabular*}{\textwidth}{@{\extracolsep{\fill}} l cccc c}
\toprule
\textbf{Model} & \textbf{Significance Level} ($\alpha$) & \textbf{Threshold} $T_\alpha$ & \textbf{Detection Score} $S$ & \textbf{Margin} ($S - T_\alpha$) & \textbf{Detected} \\
\midrule

\multirow{5}{*}{\textbf{Mistral-7B}} 
& $1\mathrm{e}{-02}$ & 1.5838 & 3.8438 & 2.2600 & \checkmark \\
& $1\mathrm{e}{-03}$ & 2.1762 & 3.8438 & 1.6675 & \checkmark \\
& $1\mathrm{e}{-04}$ & 2.6639 & 3.8438 & 1.1798 & \checkmark \\
& $1\mathrm{e}{-06}$ & 3.4662 & 3.8438 & 0.3776 & \checkmark \\
& $1\mathrm{e}{-08}$ & 4.1321 & 3.8438 & -0.2883 & $\times$ \\
\midrule

\multirow{5}{*}{\textbf{DeepSeek-7B}} 
& $1\mathrm{e}{-02}$ & 0.2358 & 3.7500 & 3.5142 & \checkmark \\
& $1\mathrm{e}{-03}$ & 0.3417 & 3.7500 & 3.4083 & \checkmark \\
& $1\mathrm{e}{-04}$ & 0.4288 & 3.7500 & 3.3212 & \checkmark \\
& $1\mathrm{e}{-06}$ & 0.5723 & 3.7500 & 3.1777 & \checkmark \\
& $1\mathrm{e}{-08}$ & 0.6913 & 3.7500 & 3.0587 & \checkmark \\
\bottomrule
\end{tabular*}
\caption{Detection Thresholds and Significance Test Results across Different Significance Levels $\alpha$. \texttt{FSW} successfully rejects the null hypothesis $H_0$ with a substantial positive margin even under extreme significance constraints (e.g., $\alpha = 10^{-8}$ for DeepSeek).}
\label{tab:2}
\end{table*}

\begin{table*}[t]
\centering
\small
\setlength{\tabcolsep}{8pt} 
\begin{tabular*}{\textwidth}{@{\extracolsep{\fill}} ll cccc ccc}
\toprule
\multirow{2.5}{*}{\textbf{Dataset}} & \multirow{2.5}{*}{\textbf{Method}} & \multicolumn{4}{c}{\textbf{Functional Preservation}} & \multicolumn{3}{c}{\textbf{Detectability}} \\
\cmidrule(lr){3-6} \cmidrule(lr){7-9}
& & PPL $\downarrow$ & $\Delta$PPL $\downarrow$ & HellaSwag $\uparrow$ & ARC-E $\uparrow$ & Det. Score $\uparrow$ & Bit Acc $\uparrow$ & AUC $\uparrow$ \\
\midrule

\multirow{2}{*}{\textbf{WikiText-2}} 
& Clean FT & 5.83 & 0.00 & 68.60 & 79.60 & -0.03 & -- & -- \\
& \textbf{FSW (Ours)} & 6.20 & +0.37 & 67.80 & 75.80 & \textbf{4.00} & 100\% & 0.899 \\
\midrule

\multirow{2}{*}{\textbf{C4}} 
& Clean FT & 8.69 & 0.00 & 79.84 & 77.74 & -0.01 & -- & -- \\
& \textbf{FSW (Ours)} & 9.59 & +0.90 & 77.38 & 69.28 & \textbf{4.88} & 100\% & 1.000 \\
\bottomrule
\end{tabular*}
\caption{Generalization across different training datasets on Meta-LLaMA-3-8B. The results demonstrate that \texttt{FSW} maintains consistent functional utility and high detectability regardless of the fine-tuning corpus used.}
\label{tab:dataset_generalization}
\end{table*}

\begin{table*}[t]
\centering
\small
\setlength{\tabcolsep}{6pt} 
\begin{tabular*}{\textwidth}{@{\extracolsep{\fill}} l c c ccc ccc c}
\toprule
\multirow{2.5}{*}{\textbf{Model}} & \multirow{2.5}{*}{\textbf{Candidates}} & \multirow{2.5}{*}{\textbf{Opt. $k$}} & \multicolumn{3}{c}{\textbf{Pre-Attack (Original)}} & \multicolumn{3}{c}{\textbf{Post-Attack (Robustness)}} & \multirow{2.5}{*}{\textbf{Ret. (\%)}} \\
\cmidrule(lr){4-6} \cmidrule(lr){7-9}
& & & PPL $\downarrow$ & Score $\uparrow$ & Acc $\uparrow$ & PPL $\downarrow$ & Score $\uparrow$ & Acc $\uparrow$ & \\
\midrule

LLaMA-2-7B-hf & $\{8,16,32, \dots\}$ & 16 & 5.91 & 6.09 & 100.0\% & 5.91 & 5.84 & 87.5\% & \textbf{95.90} \\

LLaMA-3-8B & $\{8,16,32, \dots\}$ & 32 & 6.20 & 4.00 & 100.0\% & 6.12 & 3.17 & 62.5\% & \textbf{79.30} \\

Qwen2.5-7B & $\{8,16,32, \dots\}$ & 32 & 7.24 & 5.09 & 100.0\% & 7.66 & 3.14 & 75.0\% & \textbf{61.66} \\

Mistral-7B-v0.3 & $\{8,16,32, \dots\}$ & 64 & 17.25 & 3.84 & 100.0\% & 6.59 & 3.23 & 75.0\% & \textbf{84.15} \\

DeepSeek-7B-Chat & $\{8,16,32, \dots\}$ & 16 & 77.31 & 3.75 & 100.0\% & 61.39 & 1.66 & 100.0\% & \textbf{44.17} \\

\bottomrule
\end{tabular*}
\caption{Optimal embedding dimension $k$ selection across backbone models. The optimal $k$ is determined by maximizing the retention rate (Ret.\%) of detection scores after robustness attacks, ensuring an effective trade-off between watermark capacity and structural stability.}
\label{tab:best_k}
\end{table*}

\label{appendix c}
\begin{table*}[t]
\centering
\small
\setlength{\tabcolsep}{8pt} 
\begin{tabular*}{\textwidth}{@{\extracolsep{\fill}} lcccc}
\toprule
\textbf{Training Phase (Qwen2.5-7B)} & \textbf{Time (s)} & \textbf{Steps} & \textbf{GPU Mem (MB)} & \textbf{CPU Mem (MB)} \\
\midrule
Standard FT (Baseline) & 1101.55 & 7500 & 29820.63 & 6111.39 \\
\midrule
Phase 1: Fisher Estimation & 76.61 & 250 & 29306.54 & 20815.56 \\
Phase 2: Inv. Estimation & 11.66 & 250 & 29287.93 & 20815.56 \\
Phase 3: Subspace Construction & 3.02 & 1 & 29268.78 & 20815.56 \\
Phase 4: Watermark Embedding & 1764.38 & 7500 & 56410.34 & 20815.56 \\
\midrule
\textbf{Total FSW} & \textbf{1855.68} & \textbf{8001} & \textbf{56410.34} & \textbf{20815.56} \\
\bottomrule
\end{tabular*}
\caption{Training overhead of \texttt{FSW} compared to standard fine-tuning. While embedding introduces additional memory requirements due to dual-model consistency constraints, the overall time overhead remains within a manageable range for 7B-scale models.}
\label{tab:overhead}
\end{table*}

\begin{table*}[t]
\centering
\small
\setlength{\tabcolsep}{8pt} 
\begin{tabular*}{\textwidth}{@{\extracolsep{\fill}} lcccc}
\toprule
\textbf{Model} & \textbf{Parameters} & \textbf{Hidden Size} & \textbf{Layers} & \textbf{Watermark Layer} ($\ell$) \\
\midrule
GPT-2 & 124M & 768 & 12 & 8 \\
LLaMA-2-7B-hf & 7B & 4096 & 32 & 16 \\
Meta-LLaMA-3-8B & 8B & 4096 & 32 & 16 \\
Qwen2.5-7B & 7B & 3584 & 28 & 14 \\
Mistral-7B-v0.3 & 7B & 4096 & 32 & 16 \\
DeepSeek-7B-Chat & 7B & 4096 & 30 & 15 \\
\bottomrule
\end{tabular*}
\caption{Architectural specifications of the evaluated language models. The watermark is consistently embedded in the middle layer ($\ell \approx L/2$) for each architecture to balance representation stability and expressiveness.}
\label{tab:model_specs}
\end{table*}

\end{document}